# Measurement of pressure effects on the magnetic and the magnetocaloric properties of the intermetallic compounds $R(Co_{1-x}Si_x)_2$ [R=Dy and Er]

Niraj K. Singh, Pramod Kumar, K. G. Suresh*

Indian Institute of Technology Bombay, Mumbai- 400076, India

A. K. Nigam

Tata Institute of Fundamental Research, Homi Bhabha Road, Mumbai- 400005, India

A. A. Coelho and S. Gama

Instituto de Física "Gleb Wataghin," Universidade Estadual de Campinas-UNICAMP, C.P. 6165, Campinas 13 083 970, SP, Brazil

*Abstract*

The effect of external pressure on the magnetic properties and magnetocaloric effect of $R(Co_{1-x}Si_x)_2$ [R= Er, Dy and x=0, 0.025 and 0.05] compounds has been studied. The ordering temperatures of both the parent as well as the Si substituted compounds are found to decrease with pressure. In all the compounds, the critical field for metamagnetic transition increases with pressure. It is seen that the magnetocaloric effect in the parent compounds is almost insensitive to pressure, while there is considerable enhancement in the case of Si substituted compounds. Spin fluctuations arising due to the magnetovolume effect play a crucial role in determining the pressure dependence of magnetocaloric effect in these compounds. Analysis of the magnetization data using the Landau theory has shown that the magnitude of the Landau coefficient ($C_3$) decreases with Si concentration whereas it is found to increase with pressure. The isothermal magnetic entropy change is found to behave in the same manner as $C_3$, both with Si concentration (at ambient pressure) as well as with the applied pressure.

*Corresponding author (email: suresh@phy.iitb.ac.in)

## 1. Introduction

The magnetocaloric effect (MCE) is an intrinsic property of all magnetic materials and is induced via the coupling of magnetic sublattice with applied magnetic field. MCE is measured in terms of isothermal magnetic entropy change and/or adiabatic temperature change. Though the MCE can be measured directly employing calorimetric techniques, indirect calculation of MCE using the magnetization and/or heat capacity data is more prevalent[1]. Magnetocaloric effect is being exploited in magnetic refrigeration technology and there is a strong demand for materials with giant MCE to be used as magnetic refrigerants. The possibility of use of magnetic refrigeration in both near room temperature as well as in the cryogenic temperature regime has led to an intense research in the field of MCE [2-8].

The variety of the magnetic properties exhibited by the rare earth (R) - transition metal (TM) intermetallic compounds in general and the occurrence of giant MCE in materials such as $Gd_5(Si,Ge)_4$ in particular have made R-TM compounds the natural probe for the fundamental studies as well as for applications based on MCE[2,5-9]. It has been established that giant MCE is possible only in materials which exhibit first order transitions (FOT), metamagnetic transitions or field-induced magneto-structural transitions. Several studies have shown that the giant MCE in $Gd_5(Si,Ge)_4$ compounds is a result of magneto-structural transitions[2,3]. Among the various R-TM intermetallics, $La(Fe,Si)_{13}$ compounds are known to posses large MCE near room temperature by virtue of FOT caused by the occurrence of itinerant electron metamagnetism (IEM) in the Fe subalttice[7]. Another class of compounds which exhibits IEM and therefore FOT are $RCo_2$ with R=Er, Ho and Dy[10-12]. The creation of the Co moment by the molecular field of R, as these compounds are cooled through their ordering temperatures ($T_C$) is termed as itinerant electron metamagnetism which leads to FOT at $T_C$. The presence of IEM in these compounds leads to significant MCE and has made these materials attractive from the point of view of magnetic refrigeration applications. Considerable effort, both in the experimental as well as theoretical fronts, has been put to understand the origin of IEM and giant MCE in these compounds[10,13-18]. According to previous theories, the critical parameter that governs the IEM and thereby FOT in the $RCo_2$ compounds is the molecular field or the ordering temperature[14]. Though these theories explained FOT in the compounds with R= Er, Ho, Dy, the second order transition observed in other compounds could not be explained satisfactorily. Recently, these models have been modified by incorporating the contributions from the spin fluctuations and magnetovolume effect[10]. The important features in the modified model are the role of the lattice parameter and the spin fluctuations in determining the order of magnetic transitions. Therefore, it is of great interest to study the magnetic and magnetocaloric properties of these compounds as a function of applied pressure. The compounds selected for the study are $DyCo_2$ and $Er(Co_{1-x}Si_x)_2$ [x= 0, 0.025 and 0.05]. Though the pressure dependence of magnetic and electrical properties have been studied in $RCo_2$ compounds[19-23], to the best of our knowledge this is the first report on the influence of pressure on the MCE in parent as well as substituted $RCo_2$ compounds. Furthermore, probably for the first time, a thermodynamic analysis based on the Landau theory of phase transitions has been carried out to study the change in the order of magnetic transition in these compounds under nonmagnetic substitution as well as applied pressure.

## 2. Experimental details

The preparation and the characterization techniques for all the compounds have been reported elsewhere[12,24]. The magnetization (M) measurements under various applied pressures (P) have been performed using a Cu-Be clamp type cell, which can work up to 12 kbar, attached to a SQUID magnetometer. The magnetic ordering temperatures have been calculated from the $(dM/dT)$ plots obtained from the 'field-cooled' (FC) magnetization variation as a function of temperature (T). In the FC mode the samples were cooled in presence of a field (H) of 200 Oe and the magnetization was measured under the same field in the warming cycle. The MCE of all the compounds have been determined from the M-H isotherms collected in intervals of 4 K close to the ordering temperatures.

## 3. Results and discussion

The Rietveld refinement of the room temperature powder x-ray diffractograms confirms that $R(Co_{1-x}Si_x)_2$ [R= Er and Dy; and x= 0, 0.025 and 0.05] compounds are single phase and crystallize in $MgCu_2$-type (Space group Fd3m) cubic Laves phase structure. The lattice parameter (*a*) of $Er(Co_{1-x}Si_x)_2$ compounds was found to increase from 7.135 ± 0.004 Å to 7.148 ±0.005 Å as the Si concentration increases from 0 to 0.05, whereas the '*a*' value for $DyCo_2$ was estimated to be 7.179± 0.003 Å. Figure 1 shows the temperature dependence of the 'field-cooled' magnetization data of $DyCo_2$ obtained under different pressures and in an applied field of 200 Oe. It can be seen from the figure that the FOT in $DyCo_2$ is characterized by a sharp change in the magnetization near $T_C$. The variation of $T_C$ as a function of pressure is shown in the inset of Fig.1a. The M-T plots in the same field and under various pressures have been obtained in $Er(Co_{1-x}Si_x)_2$ compounds as well. Fig 1b shows the $T_C$ *vs* P plot of $Er(Co_{1-x}Si_x)_2$ compounds. It may be noticed from Fig. 1b that the $T_C$ values of the $Er(Co_{1-x}Si_x)_2$ increase with increasing Si concentration, whereas it decreases with pressure. The $dT_C/dP$ value obtained for $ErCo_2$ in the present case is in good agreement with the value reported for single crystals of $ErCo_2$[19]. The increase in the $T_C$ with Si is attributed to the magnetovolume effect resulting from the increased lattice parameter of Si substituted compounds. In this context, it may be mentioned here that, according to Khmelevskyi and Mohn[10], the condition of IEM in the $RCo_2$ compounds is satisfied only for the compounds having the lattice parameter (*a*) in the range of 7.05 Å to 7.22 Å. The compounds with *a*<7.05 Å possess nonmagnetic Co sublattice whereas in compounds with *a*> 7.22 Å, the Co sublattice is magnetic. Therefore, the increase in the lattice parameter as a result of Si substitution takes the system closer to the critical lattice parameter required for the stable moment formation in the Co sublattice, thereby contributing to the enhancement of $T_C$. A similar effect has been observed in Si substituted $DyCo_2$ and $HoCo_2$ compounds as well[12,25]. It is of interest to note that, based on the studies of the magnetic properties of $R(Co_{1-x}Al_x)_2$, Duc *et al.*[13] have also pointed out a strong volume dependence of the Co magnetic state. On the other hand, application of external pressure would cause a reduction in the lattice parameter and therefore would lead to a decrease in the $T_C$ of both the parent as well as the substituted compounds, as seen in Fig 1. It may be mentioned here that the role of magnetovolume effect on the Co magnetic state of the $RCo_2$ (R=Er, Ho) compounds has been illustrated by the pressure dependent study of magnetic properties and electrical resistivity[19,22]. Based on the resistivity studies as a function of pressure, Syshchenko *et al.*[22] have arrived at a critical pressure of ~40 kbar for the disappearance of magnetic moment in Co sublattice in $ErCo_2$. Above this critical pressure the Co sublattice becomes similar to that of Ni sublattice (which is nonmagnetic) in $ErNi_2$ and the ordering temperature becomes almost insensitive to pressure. The variation of $T_C$ in $DyCo_2$ and $Er(Co_{1-x}Si_x)_2$ compounds with pressure is shown in Table 1.

Figure 2 shows the field dependence of magnetization isotherms of $ErCo_2$ ($T_C$=35 K) collected at T= 34 K and under various pressures. The M-H plots of $Er(Co_{0.95}Si_{0.05})_2$ ($T_C$=55 K) obtained at T=58 K and at different pressures are given as an inset in this figure. It can be seen from the figure that M-H isotherm at the ambient pressure in $ErCo_2$ does not show any metamagnetic transition (IEM) whereas a clear IEM is seen in the M-H data at higher pressures. Furthermore, the critical field required for IEM increases with pressure and eventually the IEM vanishes completely for pressures as high as 7.4 kbar. A similar effect of pressure on the critical field is observed in the Si substituted compounds as well (inset of Fig. 2). Yamada has also reported a similar effect of pressure on the critical field required for IEM in $YCo_2$ compounds[26].

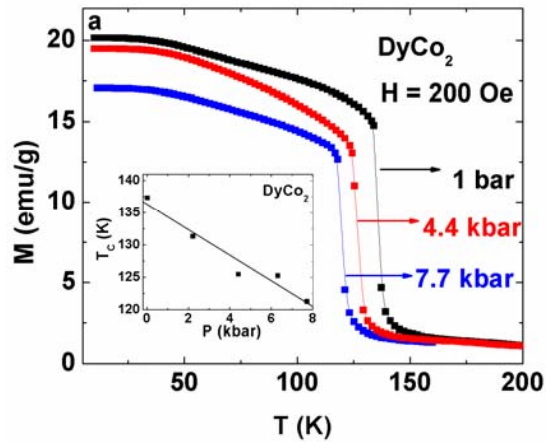

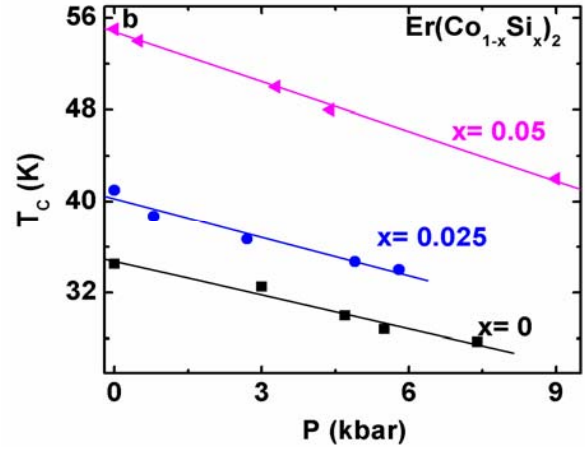

Fig. 1. (a) Temperature (T) variation of magnetization (M) data of $DyCo_2$ obtained under different pressure (P) conditions and in an applied field (H) of 200 Oe. The inset of Fig. 1a shows the pressure dependence of ordering temperatures ($T_C$) of $DyCo_2$. (b) $T_C$ vs P plot of $Er(Co_{1-x}Si_x)_2$ [x= 0, 0.025 and 0.05] compounds.

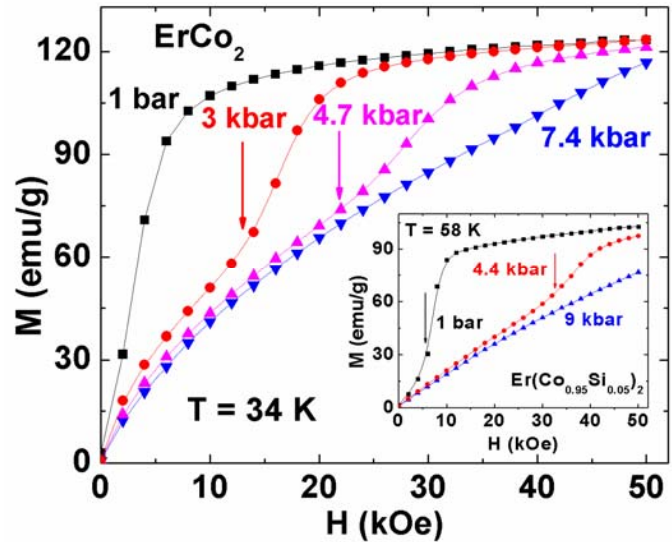

Fig. 2. Field (H) dependence of the magnetization (M) isotherm of $ErCo_2$, obtained at T= 34 K under various applied pressures (P). The inset of the figure shows the M vs H plot of $Er(Co_{0.95}Si_{0.05})_2$ collected at T=58 K under different pressures. The arrows in the figure indicate the onset of metamagnetic transition.

Table 1. The ordering temperature ($T_C$) and the maximum value of the isothermal magnetic entropy change ($\Delta S_M^{max}$) of $R(Co_{1-x}Si_x)_2$ [R=Er and Dy] compounds under various pressures (P).

| DyCo$_2$ | | | ErCo$_2$ | | | Er(Co$_{0.975}$Si$_{0.025}$)$_2$ | | | Er(Co$_{0.95}$Si$_{0.05}$)$_2$ | | |
| --- | --- | --- | --- | --- | --- | --- | --- | --- | --- | --- | --- |
| P (kbar) | $T_C$ (K) | $\Delta S_M^{max}$ (Jkg$^{-1}$K$^{-1}$) | P (kbar) | $T_C$ (K) | $\Delta S_M^{max}$ (Jkg$^{-1}$K$^{-1}$) | P (kbar) | $T_C$ (K) | $\Delta S_M^{max}$ (Jkg$^{-1}$K$^{-1}$) | P (kbar) | $T_C$ (K) | $\Delta S_M^{max}$ (Jkg$^{-1}$K$^{-1}$) |
| 0 | 137 | 12 | 0 | 35 | 33 | 0 | 41 | 27.4 | 0 | 55 | 22.7 |
| 4.4 | 126 | 12.3 | 4.7 | 30 | 32.8 | 2.7 | 37 | 28.8 | 3.3 | 50 | 24.8 |
| 7.7 | 121 | 12.4 | 7.4 | 28 | 32.5 | 5.8 | 34 | 29.9 | 9 | 42 | 26.6 |

The magnetocaloric effect, in terms of isothermal entropy change ($\Delta S_M$), for all the compounds has been calculated using the Maxwell's relation[12]. Fig. 3 shows the $\Delta S_M$ vs T plot of $Er(Co_{1-x}Si_x)_2$ compounds at ambient pressure. It can be seen from the figure that the $\Delta S_M$ vs T plot of all the compounds show a maximum near the $T_C$. The maximum value of $\Delta S_M$ ($\Delta S_M^{max}$) for ErCo$_2$, for $\Delta H$ =50 kOe is 33 Jkg$^{-1}$K$^{-1}$. The $\Delta S_M^{max}$ of Er(Co$_{0.95}$Si$_{0.05}$)$_2$, for the same $\Delta H$ is found to be 22.7 Jkg$^{-1}$K$^{-1}$. The decrease in $\Delta S_M^{max}$ with increase in Si concentration is consistent with previous reports[11,24,27]. A similar reduction in the MCE values has been observed in Dy(Co,Si)$_2$ compounds also[12]. It has been reported that the partial substitution of Co by Si in DyCo$_2$ leads to an increase in the lattice parameters, thereby leading to the enhancement of spin fluctuations. The presence of spin fluctuations decreases the strength of IEM and consequently weakens the first order nature of the transition at $T_C$. Therefore, the MCE decreases with increase in Si concentration. The detrimental role of the spin fluctuations on the strength of IEM and MCE has been reported by Han *et al.* as well[28].

Table 1 gives the summary on the pressure dependence of $T_C$ and $\Delta S_M^{max}$ in $DyCo_2$ and $Er(Co_{1-x}Si_x)_2$ compounds. Fig. 4a &b show the $\Delta S_M$ *vs* T plots of $ErCo_2$ and $Er(Co_{0.95}Si_{0.05})_2$, calculated for various applied pressures. It can be seen from this figure (also from Table 1) that, with increase in pressure, the peak in the $\Delta S_M$ *vs* T plot moves towards low temperatures in both the compounds. However, in the former case, the $\Delta S_M^{max}$ value almost remains insensitive to pressure while in the latter case $\Delta S_M^{max}$ is found to increase. It is evident from Table 1 that the pressure dependence of MCE in $DyCo_2$ is similar to that of $ErCo_2$, while the dependence in $Er(Co_{0.975}Si_{0.025})_2$ is similar to that of $Er(Co_{0.95}Si_{0.05})_2$. The insensitiveness of MCE on applied pressure in the case of $ErCo_2$ and $DyCo_2$ may be due to the fact that the strength of IEM has diminished only nominally even at a pressure of about 7.7 kbar. In fact Hauser *et al.*[21] have reported that the discontinuity (at $T_C$) in the magnetic contribution to the electrical resistivity in $ErCo_2$ decreases to about 60 % as the pressure is increased from 1 kbar to ~16 kbar, which is attributed to the reduction in the strength of IEM. In view of this, it is reasonable to assume that for a pressure of 7.7 kbar, the reduction in the IEM strength is not very much and therefore would contribute only to a nominal reduction in $\Delta S_M^{max}$. However, the reduction in $T_C$ brought about by pressure would try to increase $\Delta S_M^{max}$ due to the reduction in the thermal spin fluctuations. Therefore, it is quite possible that the reduction in MCE caused by the weakening of IEM is just compensated by the increase in MCE arising out of the reduction in $T_C$. Though the pressure dependence of MCE in $DyCo_2$ is similar to that of $ErCo_2$, the scenario may be slightly different in the former. By virtue of the larger lattice parameter, the pressure dependence of the strength of IEM would be weaker in $DyCo_2$[21]. Therefore, the insensitiveness of MCE on pressure seen in the case of $DyCo_2$ is consistent with the observations made by Hauser et al.[21]

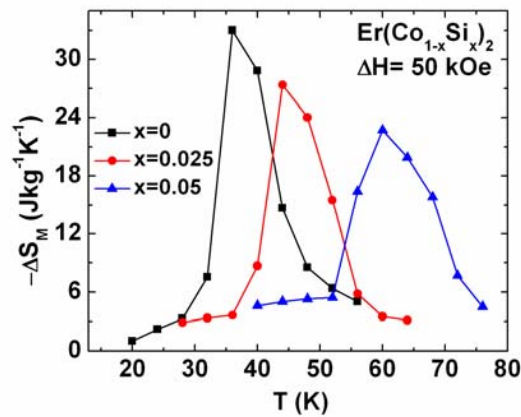

Fig. 3 Temperature (T) variation of isothermal entropy change ($\Delta S_M$) of $Er(Co_{1-x}Si_x)_2$ compounds at ambient pressure for a field change ($\Delta H$) of 50 kOe.

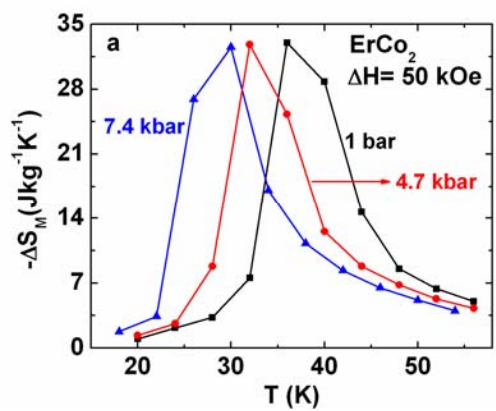

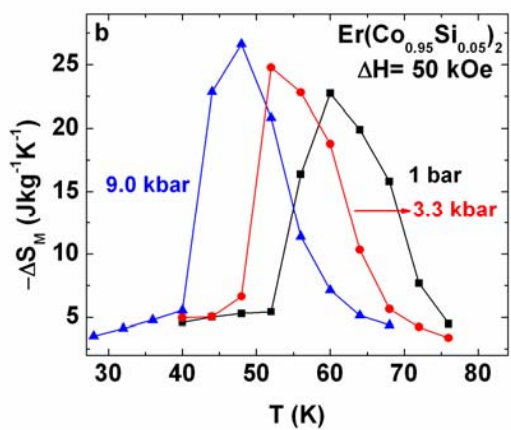

Fig. 4a & b $\Delta S_M$ *vs* T plots of $ErCo_2$ and $Er(Co_{0.95}Si_{0.05})_2$ under various pressures. All the $\Delta S_M$ values have been calculated for a field change ($\Delta H$) of 50 kOe.

On the other hand, the pressure dependence of $\Delta S_M^{max}$ in the case of Si substituted compounds is quite considerable. The increase seen in $\Delta S_M^{max}$ may be attributed to the increase in the strength of IEM, which results from the reduction of spin fluctuations related to the magnetovolume effect. In order to understand the effect of pressure on the Si substituted compounds, we have studied the nature of magnetic transition occurring in these compounds as a function of pressure. This has been done by calculating the temperature variation of the Landau coefficients. It is well known that the magnetic free energy, *F(M,T)*, in general can be expressed as Landau expansion in the magnetization as:

$$F(M,T) = \frac{C_1}{2}M^2 + \frac{C_3}{4}M^4 + \frac{C_5}{6}M^6 + ...... - \mu_0 HM \quad .......(1)$$

where $C_1$, $C_3$ and $C_5$ are the Landau coefficients which are temperature dependent. The temperature and magnetic field dependences of *F(M,T)* determine the nature of the magnetic transition. The Landau coefficients can be calculated using the equation of state, given by:

$$\mu_0 H = C_1 M + C_3 M^3 + C_5 M^5 \quad ..........(2)$$

It may be noted from equation 2 that the magnetization isotherms obtained at various temperatures may allow one to determine the temperature variation of the Landau coefficients. It is well known that the temperature dependence of the Landau coefficients may be utilized to distinguish between the first and second order transitions of magnetic materials[7,15,29]. Generally, the compounds with FOT possess positive values for $C_1(T_C)$, $C_5(T_C)$ and negative value for $C_3(T_C)$. Furthermore it has been reported that the magnitude of the $C_3$ at temperatures well below $T_C$ determines the magnitude of MCE in giant magnetocaloric materials[16].

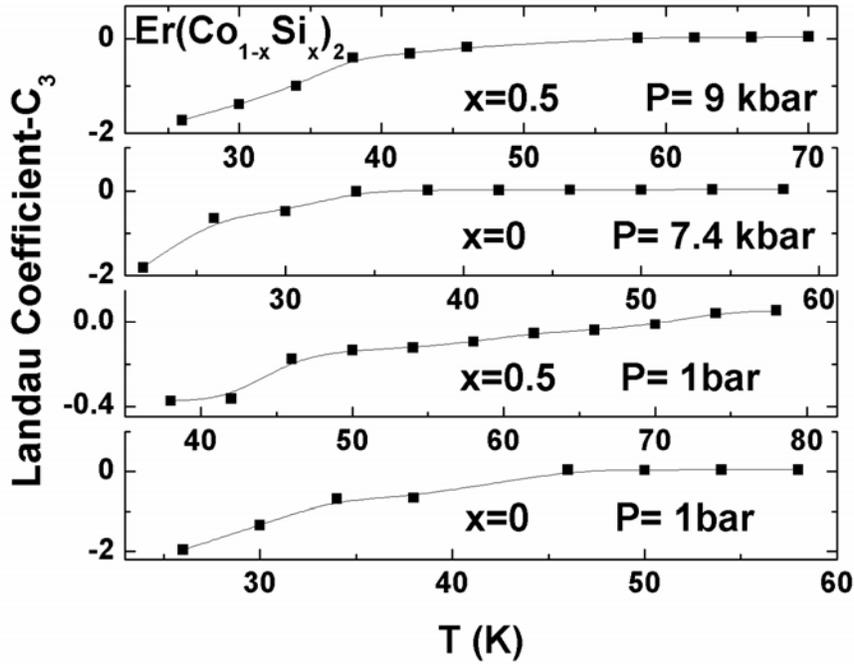

Fig. 5 Temperature (T) dependence of the Landau coefficient $C_3$, obtained under various external pressures (P), in $Er(Co_{1-x}Si_x)_2$ compounds with x=0, 0.05. [$C_3$ values have been calculated in cgs units].

The Landau coefficients of all the compounds have been determined from the M-H isotherms obtained at various temperatures. The temperature variation of $C_1$ of all the compounds exhibits a minimum near their $T_C$. Figure 5 shows the temperature variation of the coefficient $C_3$ of $Er(Co_{1-x}Si_x)_2$ [with x=0 and 0.05] under different pressures. It may be noted from the figure that for both the compounds the sign of $C_3$ near $T_C$ is negative and that its magnitude decreases with increase in temperature. A similar trend has been obtained for the $Er(Co_{0.975}Si_{0.025})_2$ and $DyCo_2$ as well. Therefore the temperature variation of $C_3$ of all the compounds indicates the presence of FOT at $T_C$. With increase in the temperature, the sign of $C_3$ of all the compounds changes from negative to positive, thereby ruling out the possibility of FOT after a limited temperature range above $T_C$. It may also be seen from Fig. 5 that at ambient pressure, the magnitude of $C_3$ in $Er(Co_{0.95}Si_{0.05})_2$ at temperatures well below $T_C$ is lower than that of $ErCo_2$. The low temperature $C_3$ value of $Er(Co_{0.95}Si_{0.05})_2$ increases with increasing pressure whereas no significant change in the $C_3$ value has been observed in $ErCo_2$. In the case of $DyCo_2$, the variation was similar to that of ErCo$_2$. This implies that the strength of IEM increases considerably in the Si substituted compounds whereas there is no significant change in the parent compounds. It may also be noticed from Fig. 5 that the difference between the low temperature C$_3$ values of ErCo$_2$ and the Si-substituted compounds almost vanishes with increase in pressure. At this point it is worth noting the fact that the $\Delta S_M^{max}$ value in the case of ErCo$_2$ and DyCo$_2$ is almost insensitive to the pressure change, but $\Delta S_M^{max}$ is quite dependent on pressure in the case of Si-substituted ErCo$_2$ compounds. Therefore, the variations of the magnitude of C$_3$ with Si concentration (at ambient pressure) as well as with pressure are consistent with the MCE variation, which implies that there is a strong correlation between the C$_3$ value and the MCE in RCo$_2$-based IEM systems. In this context, it is of importance to mention that Yamada *et al.*[16] have indeed shown that the MCE in IEM systems is primarily governed by the magnitude of C$_3$. Fujita *et al.* have also reported a similar dependence of MCE on C$_3$ in La(Fe,Si)$_{13}$ compounds which also is a well known IEM system[30].

It has been mentioned above that the partial substitution of Si for Co in ErCo$_2$ decreases the magnitude of C$_3$ at ambient pressure. We attribute this decrease to the reduction in the strength of IEM which, in turn, may be assumed to arise from the magnetovolume effect. Local magnetic moments developed as a result of increased lattice parameter after Si substitution (magnetovolume effect) are not properly exchange coupled and therefore they act as spin fluctuations. The effect of these fluctuations on the magnetic and electrical resistivity behavior has already been reported in R(Co,Al)$_2$ compounds[31]. These fluctuations may suppress the IEM and hence the C$_3$ values. On the other hand, the effect of applied pressure (in the Si-substituted compounds) is to compete with the positive magnetovolume effect and to reduce it. Consequently, the magnetic nature of the Co sublattice in the Si-substituted compounds is more or less made similar to that of ErCo$_2$. This is exactly seen in Fig.5 which shows that at high pressures, the low temperature C$_3$ value of Er(Co$_{0.95}$Si$_{0.05}$)$_2$ is almost equal to that of ErCo$_2$. Unfortunately, there are no reports available on the pressure dependent electrical resistivity studies on substituted RCo$_2$ compounds, which would have been quite useful in this context.

## 4. Conclusions

In conclusion, we have studied the effect of pressure on the magnetic and magnetocaloric properties of $R(Co_{1-x}Si_x)_2$ compounds which show IEM. The application of pressure results in the reduction of the ordering temperature, both in the parent as well as in the Si-substituted compounds. The external pressure causes an enhancement in the critical field required for IEM. The magnetocaloric effect in the parent compounds is found to be almost insensitive to pressure, while there is considerable enhancement in the case of Si substituted compounds. Due to the presence of enhanced spin fluctuations (arising due to the magnetovolume effect) in the Si substituted compounds, the influence of pressure is more visible. With increase in pressure, the negative effect of pressure is found to compete with the magnetovolume effect, thereby enhancing the MCE. The isothermal magnetic entropy change is found to behave in the same manner as $C_3$, both with Si concentration (at ambient pressure) as well as with the applied pressure.

**Acknowledgments**

One of the authors (KGS) would like to thank DST, Government of India for financial support in the form of a sponsored project.

**References:**

[1] V. K. Pecharsky and K. A. Gschneidner Jr., J. Appl. Phys. **86**, 565 (1999).

[2] V. K. Pecharsky and K. A. Gschneidner Jr., Phys Rev. Lett. **78,** 4494 (1997).

[3] K. A. Gschneidner Jr., V. K. Pecharsky and A. O. Tsokol, Rep. Prog. Phys. 68, 1479 (2005).

[4] S. Gama, A. A. Coelho, A. De Campos, A. Magnus, G. Carvalho and F. C. G. Gandra, Phys Rev. Lett. **93,** (2004) 237202.

[5] O. Tegus. E. Bruck, L. Zhang, Dagula, K. H. J. Buschow, F. R. de Boer, Physica B **319**, 174 (2002).

[6] Niraj K. Singh, S. Agarwal, K. G. Suresh, R. Nirmala, A. K. Nigam and S. K. Malik, Phys. Rev. B, **72** 14452 (2005).

[7] A. Fujita, S. Fujieda, Y. Hasegawa and K. Fukamichi, Phys. Rev. B **67**, 104416 (2003).

[8] A. M. Tishin and Y. I. Spichkin, *The Magnetocaloric Effect and its Applications* (IOP, New York, 2003).

[9] K. A. Gschneidner Jr., J. Alloys and Comp. **344**, 356 (2002).

[10] S. Khmelevskyi and P. Mohn, J. Phys. Condens. Matter **12**, 9453 (2000).

[11] N. H. Duc, D. T. Kim Anh, and P. E. Brommer, Physica B **319**, 1 (2002).

[12] Niraj K. Singh, K. G. Suresh, and A. K. Nigam, Solid State Commun. **127**, 373 (2003).

[13] N. H. Duc, T. D. Hein, P. E. Brommer and J. J. M. Franse, J. Magn. Magn. Mater. **104-107**, 1252 (1992).

[14] D. Bloch, D. M. Ewards, M. Shimizu and J. Voiron, J. Phys. F. Met. Phys. **5**, 1217 (1975); J. Inoue and M. Shimizu, J. Phys. F. Met. Phys., **12**, 1811 (1982).

[15] H. Yamada, Phys. Rev. B. **47,** 11211 (1993).

[16] H. Yamada and T. Goto, Phys. Rev. B. **68,** 184417 (2003).

[17] N. A. de Oliveira, P. J. Von Ranke, M. V. T. Costa and A. Troper, Phys. Rev. B **66,** 094402 (2002)

[18] A. Giguere, M. Foldeaki, W. Schnelle and E. Gmelin, J. Phys. Condens. Matter **11,** 6969 (1999)

[19] J. Woo, Y. Jo, H. C. Kim, A. Progov, J. G. Park, H. C. Ri, A. Podlesnyak, J. Schefer, Th. Strassle and A. Teplykh, Physica B **329-333**, 653 (2003)

[20] R. Hauser, E. Bauer and E. Gratz, H. Muller, M. Rotter, H. Michor, G. Hilscher, A. S. Markosyan, K. Kamishima and T. Goto, Phys. Rev. B **61**, 1198 (2000)

[21] R. Hauser, E. Bauer and E. Gratz, Phys. Rev. B **57**, 2904 (1998)

[22] O. Syschenko, T. Fujita, V. Sechovsky, M. Divis and H. Fujii, Phys. Rev. B. **63**, 54433 (2001).

[23] T. D. Cuong, L. Havela, V. Sechovsky, A. V. Andreev, Z. Arnold, J. Kamarad and N. H. Duc, J. Appl. Phys. **8**, 4221 (1997).

[24] Niraj K. Singh, S. K. Tripathy, D. Banerjee, C. V. Tomy, K. G. Suresh, and A. K. Nigam, J. Appl. Phys. **95**, 6678 (2004).

[25] N. H. Duc and T. K. Oanh, J. Phys. Condens. Matter **9**, 1585 (1997).

[26] H. Yamada, J. Magn. Magn. Mater. **139**, 162 (1995).

[27] D. Vasylyev, J. Prokleska, J. Sebek and V. Sechovsky, J. Alloys. Compounds **394**, 96 (2005)

[28] Z. Han, Z. Hua, D. Wang, C. Zhang, B. Gu and Y Du, J. Magn. Magn. Mater. **302**, 109 (2006).

[29] E. Gratz and A. S. Markosyan, J. Phys. Condens. Matter **13**, R385 (2001).

[30] A. Fujita and K. Fukamichi, IEEE Trans on Magn, **41**, 3490 (2005).

[31] N. H. Duc, V. Sechovsky, D. T. Hung and N. H. Kim-Ngan, Physica B **179**, 111 (1992).